\documentclass{autart}

\usepackage{amsmath,amssymb,amsfonts}
\usepackage{mathrsfs}
\usepackage{algorithmic}
\usepackage{algorithm}
\usepackage{array}
\usepackage[caption=false,font=normalsize,labelfont=sf,textfont=sf]{subfig}
\usepackage{textcomp}
\usepackage{stfloats}
\usepackage{url}
\usepackage{verbatim}
\usepackage{graphicx}
\usepackage{cite}

\usepackage{xcolor}
\usepackage{layout}
\usepackage{hyperref}
\usepackage{float}
\usepackage{fancyhdr}
\usepackage{lmodern}

\definecolor{Grey}{RGB}{128,128,128}
\definecolor{Black}{RGB}{0,0,0}
\definecolor{burntSiennaRed}{RGB}{230,111,81}
\definecolor{pacificBlue}{RGB}{33,158,188}
\definecolor{violetPurple}{RGB}{186,104,200}

\newcommand{\passive}{\textsf{passive}}
\newcommand{\ideal}{\textsf{ideal}}

\newcommand{\greedyx}{\textsf{GS$_x$}}
\newcommand{\greedyxy}{\textsf{iGS}}
\newcommand{\QBC}{\textsf{QBC}}
\newcommand{\iRDM}{\textsf{iRDM}}
\newcommand{\idealsysid}{\textsf{ideal-sysid}}
\newcommand{\idealsysidnarx}{\textsf{ideal-sysid-NARX}}
\newcommand{\idealsysidss}{\textsf{ideal-sysid-SS}}

\newcommand{\smallmat}[1]{\left[ \begin{smallmatrix}#1 \end{smallmatrix} \right]}
\newcommand{\rr}{{\mathbb R}}
\newcommand{\UU}{\mathcal{U}}

\let\labelindent\relax
\usepackage{enumitem}
\renewcommand{\theenumi}{\arabic{enumi}}

\renewcommand{\theenumii}{\arabic{enumii}}

\renewcommand{\theenumiii}{\arabic{enumiii}}

\renewcommand{\theenumiv}{\arabic{enumiv}}

\newcounter{remark}
\renewcommand{\theremark}{\arabic{remark}}

\newenvironment{remark}{%
  \refstepcounter{remark}%
  \par\noindent\textbf{Remark \theremark.} \normalfont%
}{\par\vspace{1ex}}

\begin{document}

\begin{frontmatter}

\title{Online design of experiments by active learning \\ for nonlinear system identification}


\author[IMT]{Kui Xie}\ead{kui.xie@imtlucca.it},
\author[IMT]{Alberto Bemporad}\ead{alberto.bemporad@imtlucca.it}

\address[IMT]{IMT School for Advanced Studies Lucca, Lucca, Italy.} 

\thanks[footnoteinfo]{This paper was not presented at any IFAC 
meeting. 
This research was supported by the European Research Council (ERC), Advanced Research Grant COMPACT (Grant Agreement No. 101141351). 
Corresponding author K.~Xie. 
}




\begin{keyword} 
	Design of experiments, nonlinear system identification, active learning, extended Kalman filtering.
\end{keyword}

\begin{abstract}
	We investigate the use of active-learning (AL) strategies to generate the input excitation signal at runtime for system identification of linear and nonlinear autoregressive and state-space models. We adapt various existing AL approaches for static model regression to the dynamic context, coupling them with a Kalman filter to update the model parameters recursively, and also cope with the presence of input and output constraints. We show the increased sample efficiency of the proposed approaches with respect to random excitation on different nonlinear system identification benchmarks.
\end{abstract}

\end{frontmatter}

\section{Introduction}
Many system identification approaches exist, both for linear~\cite{Lju99} and nonlinear systems~\cite{LATS20,PAGLRS25}. These methods typically rely on a given training dataset to estimate the model parameters that best approximate the system's behavior. Regardless of how well the model class is chosen or how advanced the method used to solve the training problem is, the quality of the identified model ultimately depends on the richness of the information contained in the training data. Relying solely on collecting more data can be costly, may introduce excessive redundancy without significantly increasing the information content, and complicate the optimization problem required to estimate the model parameters, due to the larger number of loss terms in the objective function to minimize~\cite{MG88,LTN15,RWGF07}.

\begin{figure}
	\centering
	\includegraphics[width=1.0\columnwidth]{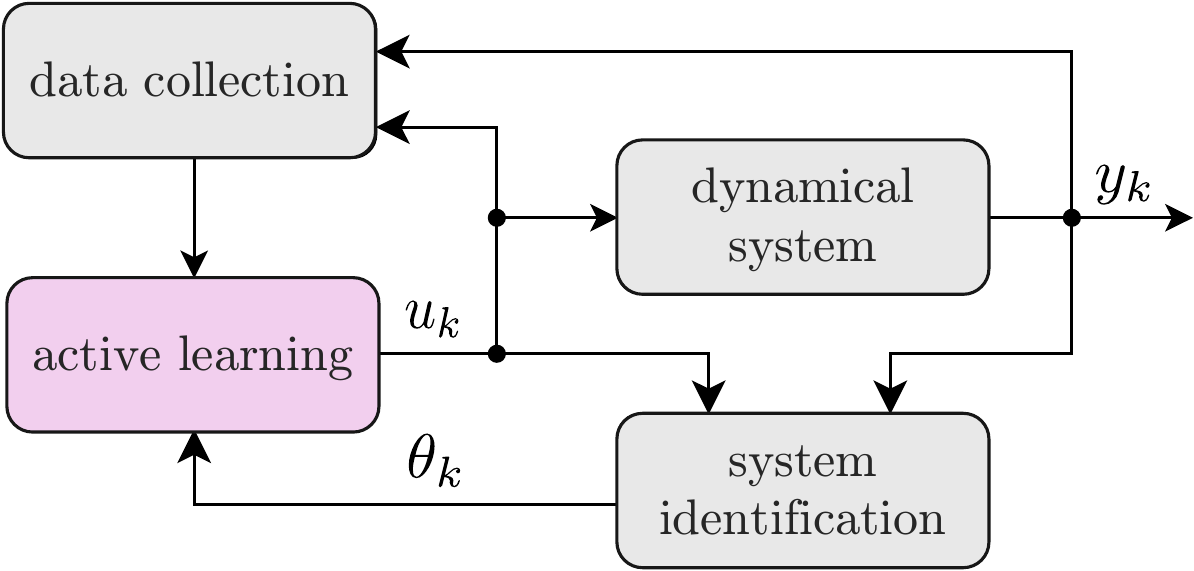}
	\caption{Online active learning for system identification.
    }
	\label{fig:ALforSI}
\end{figure}

The problem of optimal design of experiments (DoEs) has been studied for decades, dating back to the 1930s~\cite{Fis35}. In the machine learning literature, the related problem of selecting the most informative samples to query for the target value is referred to as \emph{active learning} (AL)~\cite{Set12,KG20}. AL strategies aim to reduce the number of required training samples by allowing the training algorithm to select the feature vectors to query.
Several AL methods have been proposed in the literature, primarily for classification problems~\cite{AKGHP14}, although contributions also exist for regression tasks~\cite{Mac92,CGJ96,BRK07,CZZ13,WLH19,LJLFLW21,Bem23c,IPRI24}.
Such methods mainly focus on learning {\it static} models to explain the relationship between feature vectors and targets. The samples can be arbitrarily selected from a dense set of admissible values, a pre-determined discrete pool, or a stream of feature-vector samples~\cite{Set12}. However, actively learning {\it dynamical} models is a more challenging task because not all the components of the feature vector can be changed instantaneously. Research on AL for system identification is therefore limited, and it is primarily restricted to specific classes of models such as Gaussian processes~\cite{TFM22,YZN24}, linear parameter-varying models~\cite{CMU20}, autoregressive models~\cite{XB24}, nonlinear state-space models~\cite{KTS24}, and neural-network state-space models~\cite{LRHRGG23}. Furthermore, these approaches assume that the state $x_k$ is measurable, while often only input/output data are available for system identification.

\subsection{Contribution}
We extend the AL methods reported in~\cite{YK10, WLH19, RH95, BRK07, Bem23c} for the regression of static functions to the dynamic context, focusing on learning black-box parametric models from input/output data. Specifically, we consider the problem of identifying autoregressive models, either linear (ARX) or nonlinear (NARX), and state-space models in recurrent neural network (RNN) form. A schematic diagram of the proposed strategy is shown in Fig.~\ref{fig:ALforSI}: each time a new input sample $u_k$ is generated by the AL algorithm and an output sample $y_k$ is acquired, we update the model parameters $\theta_k$ via a linear or extended Kalman filter (EKF)~\cite{Lju79,Bem23b}, depending on the chosen model class. 

The proposed recursive approach employs online optimization, based on the data collected so far, to design the experiment at runtime. As for supervised AL of static models, the developed DoE strategies ensure that the collected data are informative and diverse~\cite{Set12}, i.e., respectively, are acquired to minimize modeling errors and explore the state/action space, trying to avoid repetitions. Based on the AL method for regression proposed in~\cite{Bem23c}, we use an acquisition method based on a non-probabilistic measure of the uncertainty associated with output predictions to sample the system where uncertainty is expected to be most significant, and employ inverse-distance weighting (IDW) functions to ensure the exploration of areas not visited before. Related online AL algorithms have been proposed recently for improving the sample efficiency of reinforcement learning (RL)~\cite{LZPM24}, model predictive coverage control~\cite{RKSZC24}, Bayesian optimization (BO)~\cite{Sur24,YLBAHKHYA25}, latent space dynamics identification~\cite{HTBC24}, covariance steering~\cite{KT24}, and deep metric learning~\cite{VYCJSS24}.

We consider both one-step-ahead AL formulations, based on the uncertainty associated with the following predicted output, and a less myopic multi-step-ahead AL approach, based on the uncertainty related to the predicted outputs over a finite horizon. Moreover, we also take into account the presence of input and output constraints, with the latter treated as soft to avoid excessive conservativeness, especially at early stages when the model is very uncertain. The online computation effort of the proposed AL algorithm is limited, in particular, when the input can be selected from a discrete set (e.g., as in the case of pseudo-random binary signal excitation, where the set has only two elements). 
However, if the sampling frequency used to collect data is too high for real-time optimization, our algorithm can still be applied to a digital twin. In this case, the input sequence generated during the simulation can be saved and later used to excite the actual process.

Compared to our conference paper~\cite{XB24}, in this paper we introduce several key improvements. Besides considering more complex NARX models, such as small-scale neural networks, we extend the AL methods to {\it state-space} models, including RNN-based models, which pose additional challenges due to the presence of hidden states. We also validate the proposed algorithms on a much broader range of practical examples, demonstrating their effectiveness compared to random excitation in a set of realistic scenarios. 


The paper is structured as follows. In Sections~\ref{sec:NARX} and~\ref{sec:ss}, we present the proposed algorithm for NARX models and state-space models, respectively. Numerical experiments on nonlinear state-space models are reported in Section~\ref{sec:num-exp}. Finally, conclusions are drawn in Section~\ref{sec:cons}.

\section{Active Learning of NARX Models}\label{sec:NARX}
We first consider the problem of identifying a strictly-causal NARX model in the form
\begin{subequations}
	\begin{eqnarray}
		\hat{y}_k&=&f(x_{k-1},\theta)\\
		x_{k-1}&=&[y_{k-1}'\ \ldots\ y_{k-n_a}'\ u_{k-1}'\ \ldots\ u_{k-n_b}']'
	\end{eqnarray}
	\label{eq:NARX}%
\end{subequations}
where $\hat{y}_k\in\rr^{n_y}$, $u_k\in\rr^{n_u}$, $n_a\geq 0$, $n_b\geq 0$, $\theta\in\rr^{n_\theta}$
is the vector of parameters to learn, and $k=-\max\{n_a,n_b\},-\max\{n_a,n_b\}+1,\ldots,0,1,\ldots$ is the sample index. For example, $f$ could be a linear model, $f(x,\theta)=\theta'x$, or a small-scale neural network with weights and bias terms collected in $\theta$. Our goal is to \emph{actively} generate control inputs $u_k$ at runtime, $k=0,1,\ldots,N-1$, to efficiently learn the parameter vector $\theta$, solving the posed system identification problem in a sample-efficient manner.

From now on, we assume that all input and output signals are properly scaled. In particular, we assume the {\it standard scaling} transformation
\begin{equation}
    \sigma(\alpha) = \frac{\alpha - \mu_\alpha}{\sigma_\alpha}
    \label{eq:scaling}
\end{equation}
where $(\mu_\alpha,\sigma_\alpha)$ denote the mean and standard deviation of the signal $\alpha$,
are applied on input and output signals, whose means and standard deviations --- denoted as $\mu_u$, $\sigma_u$ for the input and $\mu_y$, $\sigma_y$ for the output, respectively --- are assumed to be known.

In the sequel, we will denote by $\theta_k$ the model parameter vector obtained by training the model with the outputs collected up to time $k$
and inputs up to time $k-1$. As in most practical applications, we assume that the input $u_k$ is subject to the constraint $u_k \in \UU$, where $\UU$ is the set of valid inputs. For instance, the hyper-box $\UU = \{u \in \rr^{n_u}: u_{\min} \leq u \leq u_{\max}\}$ or, in alternative, a finite set $\UU = \{u^1, \ldots, u^M\}$,
$u^j\in \rr^{n_u}$, $j=1,\ldots,M$, such as $\UU = \{-1, 1\}$ in the case of scalar pseudorandom binary sequence excitation.

\subsection{One-step-ahead active learning}\label{sec:NARX-OSA}
Consider any \emph{acquisition function} $a:\rr^{n_x}\to\rr$ for solving an AL problem for regression~\cite{Set12} 
(as shown later, $a$ will change with the sample time $k$, in general), where
$n_x$ is the number of features; in this case, $n_x=n_an_y+n_bn_u$ is then number of past input and outputs
\[
    x_k\triangleq[y_{k}'\ \ldots\ y_{k-n_a+1}'\ u_k'\ u_{k-1}'\ \ldots\ u_{k-n_b+1}']'
\]
used in the NARX representation~\eqref{eq:NARX}
Ideally, given a new measurement $y_k$, we would like to choose 
\begin{equation*}
	x_{k}=\arg\max_{x}a(x). 
\end{equation*}
However, at time $k$, the only component in $x_{k}$ that can be freely chosen is the current input $u_k$, given that all the remaining components involve outputs and inputs already measured. 
Hence, we restrict the acquisition problem to
\begin{subequations}
	\begin{eqnarray}
		u_k&=&\arg\max_{u\in\UU}a(x_k(u))
		\label{eq:uk-AL}\\
		x_k(u)&\triangleq&[y_{k}'\ \ldots\ y_{k-n_a+1}'\ u'\ u_{k-1}'\ \ldots\ u_{k-n_b+1}']'
		\label{eq:x_k(u)}
	\end{eqnarray}%
\label{eq:acquisition_problem_uk}%
\end{subequations}
where $x_k:\rr^{n_u}\to\rr^{n_x}$ defines the feature vector corresponding to a given input selection.
Note that when the input $u_k$ is chosen from a finite set $\UU=\{u^1,\ldots,u^M\}$, problem~\eqref{eq:uk-AL} can be easily solved by enumeration, analogously to \emph{pool-based} active learning algorithms~\cite{Set12}.
The new input $u_k$ and measured output $y_{k+1}$ can be immediately used to update the model parameters $\theta_{k+1}$.
In this paper, we use an EKF to update $\theta_{k+1}$ or simply a linear Kalman filter in case $f(x_{k-1},\theta)=\theta'\bar f(x_{k-1})$ defines a linear ARX model.
Generally, the acquisition function $a(x)$ optimized in~\eqref{eq:uk-AL} depends on $\theta_k$ and all the past feature-vector/target samples $(x_{k-1},y_{k})$ collected up to step $k$. 


In~\cite{Bem23c}, the author proposed an AL method for regression, called \ideal, based on inverse distance weighting (IDW) functions, that we briefly review here and slightly extend for experiment design. Let us define the 
the squared inverse weighting functions\footnote{Alternatively, we can use $w_j(x) = {e^{-d^2(x,x_j)}}/{d^2(x,x_j)}$ to let the weight to decay more quickly as the distance of $x$ from $x_j$ gets larger.} $w_j:\rr^{n_x}\to\rr$
\begin{equation*}
	w_j(x) = \frac{1}{d^2(x,x_j)},\ j=0,\ldots,k
\end{equation*}
where the function $d^2(x,x_j) = \| x_j -x \|^2$ measures the squared Euclidean distance between (the scaled) vectors $x$ and $x_j$.
Let $v_j:\rr^n\to\rr$ be defined for $j=0,\ldots,k$ as
\begin{equation*}
	v_k(x)=\left\{\begin{array}{ll}
			1&\mbox{if}\ x=x_j\\
			0&\mbox{if}\ x=x_h,\ h\neq j\\
			\displaystyle{\frac{w_j(x)}{\sum_{h=1}^kw_h(x)}} & \mbox{otherwise}.\end{array}\right.
	\label{eq:v}
\end{equation*}
The acquisition function consists of two non-negative terms:
\begin{equation}
	a(x)=s^2(x)+\delta z(x)
	\label{eq:aq-AL}
\end{equation}
where the \emph{IDW variance} function 
\begin{equation}
	s^2(x)=\sum_{j=0}^kv_j(x)\|y_j-f(x_{j-1},\theta_k)\|_2^2
	\label{eq:s2}
\end{equation}
is a proxy for the variance of the output $y$ predicted by the model at $x$,
$z(x)$ is the \emph{IDW exploration} function,
\begin{equation}
	z(x) = 
	\begin{cases} 
		0 & x \in \{ x_0, \dots, x_{k} \} \\
		\frac{2}{\pi} \tan^{-1} \left( \frac{1}{\sum^{k}_{j=0} w_j(x)} \right) & \text{otherwise}
	\end{cases}
	\label{eq:z}
\end{equation}
and $\delta\geq 0$ is a trade-off coefficient between exploitation (of the model $\theta_k$, used to estimate model uncertainty) and pure exploration (since $z(x)=0$ at each $x_j$ sampled so far, $j=0,\ldots,k$).

Besides \ideal, we will consider also the following alternative incremental AL methods: the greedy method \greedyx~\cite[Algorithm 1]{YK10}, the greedy method \greedyxy~\cite[Algorithm 3]{WLH19}, and the query-by-committee method \QBC~\cite{RH95,BRK07}. We exclude the \iRDM~method~\cite{LJLFLW21} as it cannot be used for recursive model learning.
See~\cite[Section 3.4]{Bem23c} for a review of these methods for regression of static models.
See Section~\ref{sec:otherAL} below for the adaptation of \greedyx, \greedyxy, and \QBC\ to the dynamic context.

\subsection{Initialization}\label{sec:NARX-Init}
As typically done in most AL algorithms, we start by using \emph{passive} learning to gather $N_i$ initial pairs of input/output samples, $N_i\geq 0$. The simplest approach is to use random sampling, that is, to generate $u_0,\ldots,u_{N_i-1}$ randomly (cf.~\cite[Section 3.1]{Bem23c}).

\begin{remark}
It is well known that closed-loop system identification data --- where the input $u_k$ depends on past outputs --- can introduce bias in model estimation if not properly addressed. The input $u_k$ our AL method chooses by solving~\eqref{eq:acquisition_problem_uk} depends on past output measurements, due to $x_k(u)$. However, in our approach, the pure exploration term $z(x_k(u))$~\eqref{eq:z} is exactly designed to ensure sufficient excitation and diversity in the data, thereby mitigating the risk of bias commonly associated with closed-loop identification. While our current model adaptation relies on (extended) Kalman filtering, classical techniques such as instrumental variables (IV) could also be incorporated to further address potential closed-loop effects, an interesting direction for future work.


\end{remark}

\subsection{Constraints}\label{sec:NARX-const}
To attempt satisfying also \emph{output} constraints, we add a penalty in \eqref{eq:uk-AL} on the expected violation of output constraints. For instance, the satisfaction of output constraints
\begin{equation}
	y_{\rm min}\leq y\leq y_{\rm max}
	\label{eq:y-constraints}
\end{equation}
can be encouraged by introducing the penalty term
\begin{equation}
	\begin{aligned}
		p(x)=\,&\rho\sum_{i=1}^{n_y} \{\max\{\hat y_{k+1,i}(x,\theta_k)-y_{{\rm max},i} ,0\} \\
		&+\max\{y_{{\rm min},i}-\hat y_{k+1,i}(x,\theta_k),0\}\}
	\end{aligned}
	\label{eq:soft-penalty}
\end{equation}
where $\hat y_{k+1}=f(x,\theta_k)$ is the next output predicted by the current model with
parameter vector $\theta_k$, and $\rho$ is a penalty parameter, $\rho\ge0$. 
Then, we solve the following problem
\begin{equation}
	u_k = \arg\max_{u\in\UU}\ a(x_k(u))-p(x_k(u)).
	\label{eq:uk-ideal-constraint}
\end{equation}

A drawback of using the penalty term~\eqref{eq:soft-penalty} is that it does not account for model uncertainty, which might be quite large during the early phase of sampling. To address this issue, we consider the confidence interval proposed in~\cite{JK11} for IDW functions, which is defined as 
\begin{equation} 
	\hat{y}_{k+1}(x,\theta_k) \pm \kappa_{\alpha} s(x)
\end{equation}
where $s(x)$ is the square root of the IDW variance function $s^2(x)$~\eqref{eq:s2} and the scaling factor $\kappa_{\alpha}$ is set as the upper $\alpha$ sample quantile of 
\begin{equation*}
	\frac{|CV_i|}{s_{-(i-1)}(x_{i-1})},\ i=0,\ldots,k
	\label{eq:kappa}
\end{equation*}
where $\alpha$ is a constant, typically set to $90\%$, $CV_i = y_i - \hat{y}_{i}(x_{i-1},\theta_k)$
is the cross-validation error at $x_{i-1}$, 
$s_{-i}^2(x_i) = \sum_{\substack{j=0, \, j \neq i}}^{k} v_{j(i)}(x_i)(y_j -\hat{y}_{i+1}(x_i,\theta_k))$, $v_{j(i)}(x_i) = w_j(x_i)/\sum_{\substack{l=0, \, l \neq i}}^{k}w_l(x_i)$, 
and $w_l(x)$ is the weighting function.
To prevent over-shrinking the constraint set in~\eqref{eq:uk-ideal-constraint}, we impose a limit on the quantity $\kappa_{\alpha} s(x)\leq \beta({y_{{\rm max}} - y_{{\rm min}}})$,
where, in this case, we set $\beta = \frac{1}{3}$.
Finally, in~\eqref{eq:uk-ideal-constraint} we take into account the uncertainty of output predictions by replacing $p(x)$ 
with
\begin{eqnarray}
	p(x)&=&\rho\sum_{i=1}^{n_y} \{\max\{\hat y_{k+1,i}(x,\theta_k)-y_{{\rm max},i} + \kappa_{\alpha} s(x),0\} \nonumber\\
	&&\hspace*{-.5em}+\max\{y_{{\rm min},i}-\hat y_{k+1,i}(x,\theta_k) + \kappa_{\alpha} s(x),0\}\}.
	\label{eq:soft-penalty-shrunk}
\end{eqnarray}

\subsection{Alternative active-learning methods}
\label{sec:otherAL}
We review three different AL methods for regression, alternative to \ideal, slightly adapted here to generate input signals for system identification. 

\subsubsection{Greedy method \greedyx}
The method \greedyx~\cite[Algorithm~1]{YK10} selects the next sample $x_k$ by maximizing its minimum distance from the existing samples. This method is not model-based, as it only aims to fill the input space.

In analogy with~\eqref{eq:uk-ideal-constraint}, we extend \greedyx~based on the acquisition problem 
\begin{equation*}
	u_k=\arg\max_{u\in\UU} d_x(x_k(u))-p(x_k(u))
	\label{eq:uk-GSx-constraint}
\end{equation*}
where $d_x(x) = \min_{i=0}^{k} \|x - x_i\|_2^2$ is the minimum distance from existing samples.

\subsubsection{Greedy method \greedyxy} 
Given the predictor $f(x_k(u),\theta_k)$ trained on the available samples, the greedy sampling technique \greedyxy~\cite[Algorithm 3]{WLH19} selects the next input $u_k$ by solving the acquisition problem
\begin{equation*}
	u_k=\arg\max_{u\in\UU}\ d_x(x_k(u))d_y(x_k(u))-p(x_k(u))
	\label{eq:uk-iGS-constraint}
\end{equation*}
where $d_x(x)$ is the same as in \greedyx\ and $d_y(x) = \min_{i=0}^{k}\left\|\hat y_{k+1}(x,\theta_k) - y_i\right\|_2^2$ is the predicted minimum distance in the output space from existing output samples.

\subsubsection{Query-by-Committee method \QBC} \label{sec:AL-QBC} 
The Query-by-Committee (QBC) method for regression~\cite{RH95,BRK07} utilizes $K_{QBC}$ different predictors
$\theta_k^j$, $j=1,\ldots,K_{QBC}$. In AL of static models, the predictors are typically obtained by bootstrapping the acquired dataset. 
In contrast, as we acquire the samples online, we create $K_{QBC}$ different models by running $K_{QBC}$ (extended) Kalman filters in parallel and, at each time step, only update $K_{QBC}-1$ models after acquiring the new sample $y_k$ to diversify them. This adaptation of QBC aims at choosing the input $u_k$ that maximizes the variance of the estimated output prediction $\hat y_{k+1}(x_k(u),\theta_k^j)$: 
\begin{equation}
    \begin{aligned}
	u_k=&\arg \max_{u\in\UU} \sum_{j=1}^{K_{QBC}}\left\| \hat y_{k+1}(x_k(u),\theta_k^j) \right.\\
    &\left.- \frac{\sum_{h=1}^{K_{QBC}}\hat y_{k+1}(x_k(u),\theta_k^h)}{K_{QBC}}\right\|_2^2 
	-p(x_k(u)). 
    \end{aligned}
    \label{eq:uk-QBC}
\end{equation}

\begin{algorithm}[t]
	\caption{Online design of experiments for system identification of NARX models using active learning and inverse-distance based exploration (\idealsysidnarx).}
	\label{alg:AL_NARX_ideal}
	~~\textbf{Input}: Set $\UU$ of admissible inputs, number $N_i$ of passively-sampled inputs, length $N$ of the experiment to design, exploration hyperparameter $\delta \ge 0$, number $L\geq 1$ of prediction steps, possible upper and lower bounds $y_{\rm max}$, $y_{\rm min}$, penalty parameter $\rho\geq0$ on output constraint violations, number $N_e$ of epochs used to train the initial model via EKF.
	\vspace*{.1cm}\hrule\vspace*{.1cm}
	\begin{enumerate}[label*=\arabic*., ref=\theenumi{}]
		\item Generate $N_i$ samples $u_0,\ldots,u_{N_i-1}$ by passive learning (e.g., random sampling);
		\item Excite the system and collect $y_0,\ldots,y_{N_i-1}$; \label{algo:ideal:init-samp}
		\item Estimate $\theta_{N_i-1}$ by running EKF $N_e$ times; \label{algo:ideal:init-train}
		\item \textbf{For} $k = N_i,\dots,N$ \textbf{do}:
		\begin{enumerate}[label=\theenumi{}.\arabic*., ref=\theenumi{}.\arabic*]
			\item measure $y_k$; \label{algo:ideal:new-y}
			\item update $\theta_{k}$ by (extended) Kalman filtering; \label{algo:ideal:new-th}
			\item \textbf{If} $k<N$, get $u_k$ by solving problem~\eqref{eq:uk-multiStep-ideal}, 
			with penalty $p$ as in~\eqref{eq:soft-penalty} or~\eqref{eq:soft-penalty-shrunk} to handle possible 
			output constraints;\label{algo:ideal:new-u}
		\end{enumerate}
		\item \textbf{End}.\label{algo:ideal:stop}
	\end{enumerate}
	\vspace*{.1cm}\hrule\vspace*{.1cm}
	~~\textbf{Output}: Estimated parameter vector $\theta_{N}$;
	input excitation $u_0,\ldots,u_{N-1}$.
\end{algorithm}

\subsection{Multi-step prediction}\label{sec:AL-MSP}
So far, we only considered one-step-ahead predictions $\hat y_{k+1}$ to choose the next control input $u_k$. To circumvent such a possible myopic view, we can take a \emph{predictive} approach and extend the formulation~\eqref{eq:uk-ideal-constraint} to optimize a finite sequence of future inputs $U_k=[u_k'\ u_{k+1}'\ \ldots\ u_{k+L-1}']'$, 
\[
    U_k = \arg\max_{U\in\UU^L} \sum_{j=0}^{L-1} \left(a(x_{k+j}(U)) - p(x_{k+j}(U))\right)
\]
where $L\geq 1$ is the desired prediction horizon and $\UU^L\triangleq\UU\times\ldots\times\UU$. In this case, the predicted regressor vector $x_{k+j}$ entering the acquisition function $a$ contains either outputs $\hat y_{k+h}$ predicted by model $\theta_k$ ($h\geq 1$) or measured outputs $y_{k+h}$ ($h\leq 0$), and either current/future inputs $u_{k+h}$ ($h\geq 0$) to be optimized or past inputs (for $h<0$). Note that, by causality, the predicted regressor as $x_{k+j}(U)$ only depends on the first $j+1$ inputs in $U$.

In the case we use the \ideal~acquisition function, since future measured outputs $y_{k+j+1}$ are not available, replacing them by surrogates $\hat y_{k+j+1}=f(\hat x_{k+j},\theta_k)$ cause the IDW variance $s^2(\hat x_{k+j})=0$, and thus $a(\hat x_{k+j})=z(\hat x_{k+j})$, $\forall j>0$. Therefore, the multi-step-ahead active learning problem becomes:
\begin{equation}
	U_k\! = \!\arg\!\max_{U\in\UU^L} s^2\!(x_k(U))+\delta\!\!\sum_{j=0}^{L-1}\!(z(x_{k+j}(U))-p(x_{k+j}(U))).
	\label{eq:uk-multiStep-ideal}
\end{equation}
Note that~\eqref{eq:uk-multiStep-ideal} coincides with~\eqref{eq:uk-ideal-constraint} when $L=1$ 
as $x_{k}(U)=x_k(u)$.

Based on the receding-horizon mechanism used in model predictive control, after solving~\eqref{eq:uk-multiStep-ideal}, only the current input $u_k$ is applied to excite the process, while the remaining moves $u_{k+j}$ are discarded, for all $j=1,\ldots,L-1$. Then, after acquiring the new measurement $y_{k+1}$ and updating the model to get the new parameter vector $\theta_{k+1}$, 
problem~\eqref{eq:uk-multiStep-ideal} is solved again to get $U_{k+1}$, and so on.

The overall algorithm for online input design for NARX model system identification based on the \ideal~active learning approach, denoted by \idealsysidnarx, is summarized in Algorithm \ref{alg:AL_NARX_ideal}.

\subsection{Numerical complexity}\label{sec:AL-comp-NARX}
To analyze the complexity of Algorithm~\ref{alg:AL_NARX_ideal}, we assume that the model parameter vector $\theta$ is estimated using EKF, which has complexity $O(n_{\theta}^2)$ per step, mainly due to covariance matrix updates. The initial batch training via EKF run $N_e$ times at Step~\ref{algo:ideal:init-train} over $N_i$ samples contributes $O(n_{\theta}^2 N_e N_i)$ operations, and the recursive parameter estimation contributes further $O(n_{\theta}^2 (N-N_i))$ operations at Step~\ref{algo:ideal:new-th}. For \QBC, $(K_{QBC}-1)$ additional EKF instances increase the parameter estimation cost to $O(K_{QBC} n_{\theta}^2 N)$.

For pool-based sampling with $\UU=\{u^1,\ldots,u^M\}$, solving problem~\eqref{eq:uk-multiStep-ideal} requires $(N-N_i)M^L$ evaluations of the acquisition function $a(x)$, or $(N-N_i)M$ for single-step prediction ($L=1$) at Step~\ref{algo:ideal:new-u}. Each evaluation of~\eqref{eq:aq-AL} involves computing IDW weights and distances to all previous samples, with complexity $O(k)$ at iteration $k$, yielding total complexity $O(M N^2)$ for the acquisition function evaluations. Model predictions require $(N-N_i)M L$ evaluations of the NARX model~\eqref{eq:NARX}, each with cost $C_f$, and constraint handling via~\eqref{eq:soft-penalty} adds $(N-N_i)M L$ penalty function evaluations. Hence, the overall computational complexity is
$O\left(n_{\theta}^2 (N + N_e N_i) + M N^2 + M N L C_f\right)$.

\section{Active Learning of State-Space Models}\label{sec:ss}
The procedures presented in the previous section can be extended to learn 
strictly causal\footnote{The approach can be extended to non-strictly causal models 
with some modifications to handle the presence of I/O feedthrough.} 
state-space models of the form
\begin{equation}
	\begin{aligned}
		x_{k+1}&=f_x(x_k,u_k,\theta_x)\\
		y_{k}&=f_y(x_k,\theta_y)
	\end{aligned}
	\label{eq:ssmodel}
\end{equation}
where $x_k\in\rr^{n_x}$ is now the hidden state vector at time $k$.
Here, we focus on RNN models, in which $f_x$, $f_y$ are feedforward neural networks with weights
and bias terms collected in $\theta_x$, $\theta_y$, respectively, although the approach can be applied to
any parametric nonlinear state-space model. In order to actively learn model~\eqref{eq:ssmodel},
we must explore the space of state-input vectors $q_k=\begin{bmatrix}x_k', u_k'\end{bmatrix}'$.
However, as in the case of NARX models, only the component $u_k$ of $q_k$ can be decided arbitrarily at time $k$. Given that the hidden state $x_k$ is not modifiable (and even not measurable), we must consider its estimate $x_{k|k}$ at time $k$. Such an estimate depends on the past information $I_k\triangleq\{u_0,\ldots,u_{k-1}, y_0,\ldots,y_k\}$.

Consider the current time $k$ and let the current model-coefficient vectors $\theta_{x|k}$,  $\theta_{y|k}$ be estimated from $I_k$. Let $x_{0|k},\ldots,x_{k|k}$ be the sequence of hidden states estimated by reprocessing the dataset $I_k$ in accordance with~\eqref{eq:ssmodel} with $\theta_x=\theta_{x|k}$,  $\theta_y=\theta_{y|k}$; in particular, as suggested in~\cite[Section III.C]{Bem25}, we run an extended Kalman filter (or a linear Kalman filter, in case model~\eqref{eq:ssmodel} is linear) and Rauch-Tung-Striebel (RTS) smoothing~\cite[p.~268]{SS23} to get $x_{0|k}$, further refine $x_{0|k}$ by nonlinear programming (L-BFGS optimization), and run EKF to get the following states $x_{1|k},\ldots,x_{k|k}$. 
Then, similarly to~\eqref{eq:uk-AL}, we solve the acquisition problem
\begin{equation*}
	u_k=\arg\max_{u\in\UU}a\left(\begin{bmatrix}x_{k|k}\\u\end{bmatrix}\right)
	\label{eq:uk-AL-ss}
\end{equation*}
where the acquisition function $a$ depends on the active-learning method used,
as described next. 

\noindent\textbf{Greedy method \greedyx}.
The input $u_k$ is obtained by solving
\begin{equation*}
	\begin{aligned}
		u_k=\arg\max_{u\in\UU} d_x\left(\begin{bmatrix}x_{k|k}\\u\end{bmatrix}\right)
	\end{aligned}
	\label{eq:uk-ss-GSx-no}
\end{equation*}
where $d_x(q) = \min_{i = 0,...,k-1} \|q - q_i\|_2^2$ and $q_i=\begin{bmatrix}  x_{i|k}', u_i'\end{bmatrix}'$ is the vector
collecting existing state estimates and past input samples. 



\noindent\textbf{Greedy method \greedyxy}.
We select the input $u_k$ as
\begin{equation*}
    \begin{aligned}
		u_k=\arg\max_{u\in\UU} & d_x\left(\begin{bmatrix}x_{k|k}\\u\end{bmatrix}\right) \cdot \\
        &d_y\left(\begin{bmatrix}f_y(f_x(x_{k|k},u,\theta_{x|k}),\theta_{y|k})\\
			f_x( x_{k|k},u,\theta_{x|k})
		\end{bmatrix}\right)
    \end{aligned}
	\label{eq:uk-ss-iGS-no}
\end{equation*}
where $d_y(r) = \min_{i = 0,...,k-1} \| r - r_i \|_2^2$ is the minimum distance in the state and output space from predicted states and existing output samples
and $r_i=\begin{bmatrix}y_{i+1}',  x_{i+1|k}'\end{bmatrix}'$ is the vector
collecting past output measurements and state estimates.

\noindent\textbf{Method \ideal}.
Similarly to~\eqref{eq:uk-AL} the input $u_k$ is obtained by solving
\begin{equation*}
	u_k=\arg\max_{u\in\UU} s^2(q)+\delta z(q)
	\label{eq:uk-AL-ss-no}
\end{equation*}
where $z(q)$ is the IDW function~\eqref{eq:z} defined over the samples $q_i$, and the IDW variance function $s^2$ similar to~\eqref{eq:s2} is defined as
\begin{equation}
	\begin{aligned}
		s^2(q)=&\sum_{j=0}^{k-1}v_{j}(q)(\|f_y(f_x(q_x,q_u,\theta_{x|k}),\theta_{y|k}) - y_{j+1}\|_2^2\\
			   &+\alpha\| f_x(q_x,q_u,\theta_{x|k}) -  x_{j+1|k}\|_2^2).
		\label{eq:IDWvar-ss-no}
	\end{aligned}
\end{equation}
In~\eqref{eq:IDWvar-ss-no}, we split $q=[q_x',\, q_u']'$ and $v_j(q)$
are the IDW coefficient functions defined over the same samples $q_i$. 
The coefficient $\alpha>0$ trades off between emphasizing the uncertainty associated with the outputs and the one associated with the states. 

\noindent\textbf{\QBC~method}.
We use the same formulation as in~\eqref{eq:uk-QBC}, but with the predicted output $\hat y_{k+1}(q,\theta_{y|k}^j)=f_y(f_x(q_x,q_u,\theta_{x|k}^j),\theta_{y|k}^j)$. 

For the safe exploration of the output space, like in the NARX case~\eqref{eq:uk-ideal-constraint}, we can add a penalty term to the acquisition function and solve the following problem
\begin{subequations}
	\begin{equation}
		u_k = \arg\max_{u\in\UU}\ a(q)-p(q)
		\label{eq:uk-ideal-constraint-ss}
	\end{equation}
where
	\begin{equation}
		\begin{aligned}
			p(q)=\,&\rho\sum_{i=1}^{n_y} \{\max\{\hat y_{k+1,i}(q,\theta_{y|k}) -y_{{\rm max},i},0\} \\
			&+\max\{y_{{\rm min},i}-\hat y_{k+1,i}(q,\theta_{y|k}) ,0\}\} \label{eq:soft-penalty-ss}
		\end{aligned}
	\end{equation}
	\label{eq:acquisition_problem_uk_ss}%
\end{subequations}
and $\hat y_{k+1}(q,\theta_{y|k}) = f_y(f_x(q_x,q_u,\theta_{x|k}),\theta_{y|k})$ is the predicted output at the next time step $k+1$ given the current state estimate $x_{k|k}$ and the input $u$.

A multi-step ahead version of the active learning problem can be formulated for state-space models too, similarly to the NARX case, only using the IDW variance function $s^2$ for the first step and the IDW exploration function $z$ at all future steps in prediction, where future samples, $q_{k+j}=[x_{k+j|k}',\ u_{k+j}']'$, are predicted by running the model with the current coefficients $\theta_{x|k}$, $\theta_{y|k}$ in open-loop.

The overall algorithm for online input design for state-space model system identification based on the \ideal~active learning approach, denoted by \idealsysidss, is summarized in Algorithm \ref{alg:AL_SS_ideal}.
The model is initially trained by L-BFGS-B~\cite{BLNZ95} via the \texttt{jax-sysid} package~\cite{Bem25} 
and then refined recursively using EKF after each active-learning step. Algorithm \ref{alg:AL_SS_ideal} is also applied using the other acquisition functions \greedyx, \greedyxy, and \QBC.

\begin{algorithm}[t]
	\caption{Online design of experiments for system identification of state-space models using active learning and inverse-distance based exploration (\idealsysidss).}
	\label{alg:AL_SS_ideal}
	~~\textbf{Input}: Set $\UU$ of admissible inputs, number $N_i$ of passively-sampled inputs, length $N$ of the experiment to design, exploration hyperparameters $\delta,\alpha \ge 0$, number $L\geq 1$ of prediction steps, possible upper and lower bounds $y_{\rm max}$, $y_{\rm min}$, penalty parameter $\rho\geq0$ on output constraint violations, state-sequence reconstruction interval $m \geq 1$, number $N_b$ of epochs used to train the initial model via L-BFGS-B.
	\vspace*{.1cm}\hrule\vspace*{.1cm}
	\begin{enumerate}[label*=\arabic*., ref=\theenumi{}]
		\item Generate $N_i$ samples $u_0,\ldots,u_{N_i-1}$ by passive learning (e.g., random sampling)
		\item Excite the system and collect $y_0,\ldots,y_{N_i-1}$; \label{algo:ideal1:init-samp}
		\item Initialize model parameters $\overline{\theta}_{N_i-1}$ by L-BFGS-B $N_b$ times on collected training data; \label{algo:ideal1:init-train-L-BFGS-B}
		\item Run (extended) Kalman filter to refine estimate $\theta_{N_i-1}$ and initialize covariance matrices; \label{algo:ideal1:init-train-EKF}
		\item Reconstruct $\hat x_0, \hat x_1,\ldots,\hat x_{N_i}$ by EKF and RTS smoothing plus L-BFGS refinement; \label{algo:ideal1:init-refine}
		\item \textbf{For} $k = N_i,\dots,N$ \textbf{do}:
		\begin{enumerate}[label=\theenumi{}.\arabic*., ref=\theenumi{}.\arabic*]
			\item measure $y_k$; \label{algo:ideal1:new-y}
			\item update $\theta_{k}$ by (extended) Kalman filtering; \label{algo:ideal1:new-th}
			\item (only every $m$ steps) reconstruct $\hat x_0$, $\hat x_1$, $\ldots$, $\hat x_{k+1}$ by EKF and RTS smoothing + L-BFGS refinement; \label{algo:ideal1:new-x}
			\item \textbf{If} $k<N$, get $u_k$ by solving problem~\eqref{eq:acquisition_problem_uk_ss}; \label{algo:ideal1:new-u}
		\end{enumerate}
		\item \textbf{End}.\label{algo1:ideal:stop}
	\end{enumerate}
	\vspace*{.1cm}\hrule\vspace*{.1cm}
	~~\textbf{Output}: Estimated parameter vector $\theta_{N}$;
	input excitation $u_0,\ldots,u_{N-1}$.
\end{algorithm}

\subsection{Numerical complexity}
\label{sec:AL-comp-SS}
In analyzing the complexity of Algorithm~\ref{alg:AL_SS_ideal}, we assume that the model parameters $\theta_x$ and $\theta_y$ are initially estimated using $N_b$ L-BFGS-B iterations at Step~\ref{algo:ideal1:init-train-L-BFGS-B}, with complexity $O\left((n_{\theta_x} + n_{\theta_y}) N_b N_i C_{f_x,f_y} + (n_{\theta_x}+n_{\theta_y})^2 N_b\right)$, where the first term accounts for gradient computation through backpropagation and the second for L-BFGS-B parameter updates. Here, $C_{f_x,f_y}$ represents the cost of evaluating the state-space model~\eqref{eq:ssmodel}. Subsequently, refining the estimate using EKF at Step~\ref{algo:ideal1:init-train-EKF} and recursive parameter estimation at Step~\ref{algo:ideal1:new-th} contributes $O\left((n_{\theta_x} + n_{\theta_y})^2 N\right)$ operations.

A key difference from the NARX case is the state reconstruction at Step~\ref{algo:ideal1:new-x}: every $m$ iterations, all past states $\hat{x}_0, \ldots, \hat{x}_{k+1}$ must be reconstructed based on the latest model using EKF forward pass and RTS smoothing, requiring $O(n_x^2 k)$ operations at each iteration $k$, leading to a total $O\left(n_x^2 N^2/m\right)$ operation complexity.

For pool-based sampling, the acquisition function evaluation at Step~\ref{algo:ideal1:new-u} follows the same pattern as the NARX case, requiring $(N-N_i)M$ evaluations for single-step prediction ($L=1$). However, the feature vectors now include predicted states $q_k = [x_k', \ u_k']'$, and the IDW computations operate in this augmented space with complexity $O(k)$ at iteration $k$, yielding total complexity $O(M N^2)$. Model predictions require evaluating the state-space model~\eqref{eq:ssmodel}, with $(N-N_i)M L$ total evaluations, and constraint handling via~\eqref{eq:soft-penalty} adds $(N-N_i)M L$ penalty function evaluations.

The overall computational complexity is $O((n_{\theta_x} + n_{\theta_y}) N_b N_i C_{f_x,f_y} + (n_{\theta_x} + n_{\theta_y})^2 (N+N_b)+(n_x^2 N^2)/m + M N^2 + M N L C_{f_x,f_y})$. 
The state reconstruction term $O(n_x^2 N^2/m)$ may dominate for high-dimensional states, especially when $m=1$ (reconstruction at every step). This makes \idealsysidss~computationally more intensive than \idealsysidnarx, where state reconstruction is not required.



\section{Numerical Experiments}\label{sec:num-exp}
We assess the performance of the proposed active learning algorithms on both NARX neural network (NARX-NN) models, using Algorithm~\ref{alg:AL_NARX_ideal} (\idealsysidnarx), and RNN state-space models, using Algorithm~\ref{alg:AL_SS_ideal} (\idealsysidss), across three standard nonlinear system identification benchmarks: a two-tank system~\cite{MW20}, an ethylene oxidation plant~\cite{DEC16}, and an industrial robot arm~\cite{WG06}. For comparison, we also include passive learning and alternative active learning methods (\greedyx, \greedyxy) as described in Sections~\ref{sec:otherAL} and~\ref{sec:ss}. The \QBC~method has been implemented and tested, but is not reported in the experiments due to its poor performance, even when compared to passive sampling.

\subsection{Experimental Setup}
For each benchmark, noisy synthetic data are generated by numerically integrating the underlying system of nonlinear ordinary differential equations of the plant model with high accuracy. All methods use $N_i$ initial samples generated by random sampling, with standard scaling~\eqref{eq:scaling} applied to input/output data based on statistics from the initial dataset.

The main differences between NARX-NN and RNN settings are summarized as follows:


\noindent\textbf{Covariance matrices}. 
For NARX-NN models, EKF only estimates the neural network parameters. In this context, $Q_{\theta}\in\rr^{n_\theta \times n_\theta}$ and  $R\in\rr^{n_y \times n_y}$ are hyperparameters of the EKF representing the covariance matrices of process and measurement noise, respectively, and $P_k\in\rr^{n_\theta \times n_\theta}$ is the resulting covariance matrix of parameter estimation errors.
For RNN-based models, EKF estimates both the parameters and the hidden states of the RNN. The covariance matrix $\bar Q_x\in\rr^{n_x \times n_x}$ of state noise is a further hyperparameter. The resulting matrix $\bar P_k \in \mathbb{R}^{(n_x + n_{\theta_x} + n_{\theta_y}) \times (n_x + n_{\theta_x} + n_{\theta_y})}$ is the covariance matrix of the combined state and parameter estimatation errors.

\noindent\textbf{Model training}. NARX-NN parameters $\theta \in \rr^{n_{\theta}}$ are initially trained by running EKF $N_e = 50$ times over the same data. RNN parameters $\theta_x$ and $\theta_y$ are first trained by L-BFGS-B using \texttt{jax-sysid}~\cite{Bem25} and then refined using EKF. Both models are subsequently updated recursively via EKF.

\noindent\textbf{State estimation}. NARX-NN models do not require estimating hidden states, whereas for RNNs, hidden states are estimated at each step using EKF and RTS smoothing as required by the AL methods. As the model parameters $\theta_k$ changes relatively slowly, we only re-estimate the entire sequence of past hidden states $\hat x_0,\ldots, \hat x_{k}$ every $m$ steps, with $m = 10$. We found that this does not affect the performance of the AL methods significantly.

In the following examples, all neural network models share similar structures. Specifically, the first prediction model $f(x_{k-1},\theta_k)$ is the two-layer NARX neural network
\begin{equation}
    f(x_{k-1},\theta_k) = W_3\,\sigma_2\big(W_2\,\sigma_1(W_1 x_{k-1} + b_1) + b_2\big) + b_3
    \label{eq:NARX-f}
\end{equation}
where $x_{k-1} \in \mathbb{R}^{n_x}$ is the regressor vector, $n_x = n_a + n_b$, $n_a=3$, $n_b=3$, $n_u=1$, $n_y=1$, 
$W_1 \in \mathbb{R}^{n_1 \times n_x}$, $b_1 \in \mathbb{R}^{n_1}$, $W_2 \in \mathbb{R}^{n_2 \times n_1}$, $b_2 \in \mathbb{R}^{n_2}$, $W_3 \in \mathbb{R}^{n_y \times n_2}$, and $b_3 \in \mathbb{R}^{n_y}$. The activation functions $\sigma_1(\cdot)$ and $\sigma_2(\cdot)$ are elementwise $\arctan$ functions. The parameter vector $\theta_k$ collects all weights and bias terms.

The RNN model used for the identification of state-space models consists of the following state-update $f_x$ and output function $f_y$:
\begin{subequations}\label{eq:RNN-f}
\begin{align}
    f_x(x_k, u_k, \theta_x) = &W_3^x\,\sigma_2\big(W_2^x\,\sigma_1(W_1^x [x_k'\ u_k']' + b_1^x) + b_2^x\big)\nonumber\\
    & + b_3^x \label{eq:RNN-fx} \\
    f_y(x_k, \theta_y) = &W_2^y\,\sigma_3(W_1^y x_k + b_1^y) + b_2^y \label{eq:RNN-fy}
\end{align}
\end{subequations}
where $[x_k'\ u_k']' \in \mathbb{R}^{n_x + n_u}$ is the concatenated state and input vector. The matrices $W_1^x \in \mathbb{R}^{n_1^x \times (n_x + n_u)}$, $W_2^x \in \mathbb{R}^{n_2^x \times n_1^x}$, $W_3^x \in \mathbb{R}^{n_x \times n_2^x}$ and vectors $b_1^x \in \mathbb{R}^{n_1^x}$, $b_2^x \in \mathbb{R}^{n_2^x}$, $b_3^x \in \mathbb{R}^{n_x}$ are the weights and biases of the state-update network, with $\sigma_1(\cdot)$ and $\sigma_2(\cdot)$ being elementwise $\tanh$ activation functions. The output network uses $W_1^y \in \mathbb{R}^{n_1^y \times n_x}$, $W_2^y \in \mathbb{R}^{n_y \times n_1^y}$, $b_1^y \in \mathbb{R}^{n_1^y}$, $b_2^y \in \mathbb{R}^{n_y}$, and an elementwise $\tanh$ activation $\sigma_3(\cdot)$. The parameter vectors $\theta_x$ and $\theta_y$ collect all weights and biases of the state-update and output networks, respectively.

Performance is evaluated using one-step-ahead prediction ($L=1$). The overall prediction quality on training and test datasets is quantified using the root-mean-square error (RMSE), defined as 
\begin{equation*}
		\text{RMSE} = \sqrt{\frac{1}{N_{\rm max}} \sum_{k = 0}^{N_{\rm max}-1} (y_k-\hat{y}_k)^2}
\end{equation*}
where $\hat{y}_k = f(x_{k-1},\hat{\theta}_k)$ for NARX-NN models, while $\hat{y}_k = f_y(\hat{x}_k,\hat{\theta}_{y|k})$ for RNNs, and $N_{\rm max}$ is the number of samples in the set ($N_{\rm max}=N$ for training and $N_{\rm max}=N_{\rm test}$ for testing).
Model quality is also assessed by the $R^2$ coefficient of determination, defined as
\begin{equation}
	R^2 = \left(1 - \frac{\sum_{k=0}^{N_{\rm max}-1} (y_k - \hat{y}_k)^2}{\sum_{k=0}^{N_{\rm max}-1} (y_k - \bar{y})^2} \right)\times 100\%
	\label{eq:R2}
\end{equation}
where $\bar{y} = \frac{1}{N_{\rm max}} \sum_{k=0}^{N_{\rm max}-1} y_k$ is the mean of the output data.
To quantify output constraint violations~\eqref{eq:y-constraints}, we report the mean constraint violation (MCV) over $N_r$ runs:
\begin{equation}
	\text{MCV} = \frac{\sum_{j=1}^{N_{r}} \sum_{k=N_i}^{N}  \max\left(0, y_k^j - y_{\rm max}, y_{\rm min} - y_k^j\right)}{N_{r}(N - N_i)} 
	\label{eq:MCV}
\end{equation}
where $y_k^j$ is the $k$-th output of the $j$-th run. In all experiments, we set the penalty hyperparameter $\rho=10^{12}$. Note that in~\eqref{eq:MCV} we only account for constraint violation only during the active learning phase, $N_i \le k \le N$. 

The computations required to estimate NARX-NN models were performed in MATLAB 2023b on an Intel(R) Core(TM) i7-8750H CPU @ 2.20GHz with 16 GB RAM. The identification of RNN models was performed in Python 3.12 on an Intel(R) Xeon(R) W-2245 CPU @ 3.90GHz with 125 GB RAM.


\subsection{Two-tank benchmark}
We consider noisy synthetic data generated from a model of a two-tank system used as a nonlinear system identification benchmark in the System Identification Toolbox for MATLAB R2023b~\cite{MW20}. The plant model has two states (upper tank water level and lower tank water level), one output ($y$ = lower tank water level), and one input ($u$ = pump voltage). Samples are generated by numerically integrating the system of nonlinear ordinary differential equations of the plant model with high accuracy and collecting samples every $T_s = 0.5 \ s$, with initial state $x_0=\smallmat{0.0\\ 0.1}$. 
The measurement noise is assumed to be Gaussian with zero mean and standard deviation equal to $\gamma$ times the output value, i.e., $\eta_k = \gamma\, y_k\, \varepsilon_k$, where $\gamma = 0.02$, and $\varepsilon_k \sim \mathcal{N}(0,1)$.

We use pool-based sampling with $\UU = \{0,\ 0.01,\ \ldots,\ 10\}$, that is, the
inputs are generated by the online active learning algorithm in the interval $[0,10]$ with resolution $0.01$. 
When exciting the system to collect data while trying to enforce the output constraints
\begin{equation}
	0.03 \leq y_k \leq 0.08.
	\label{eq:TT-constraints}
\end{equation}

We first train a two-layer NARX neural network~\eqref{eq:NARX-f}, with $n_1 = 8$ and $n_2 = 6$
hidden neurons.
We set the~\idealsysidnarx~hyperparameters, $\delta = 100$, $N_i = 80$, and $N = 1000$ and apply Algorithm~\ref{alg:AL_NARX_ideal}. We also generate a test dataset of $N_{\rm test} = 2000$ samples by simulating the system with the same initial conditions and inputs randomly sampled from the same pool $\UU$. 

\begin{figure}
	\centering
	\includegraphics[width=1.0\columnwidth]{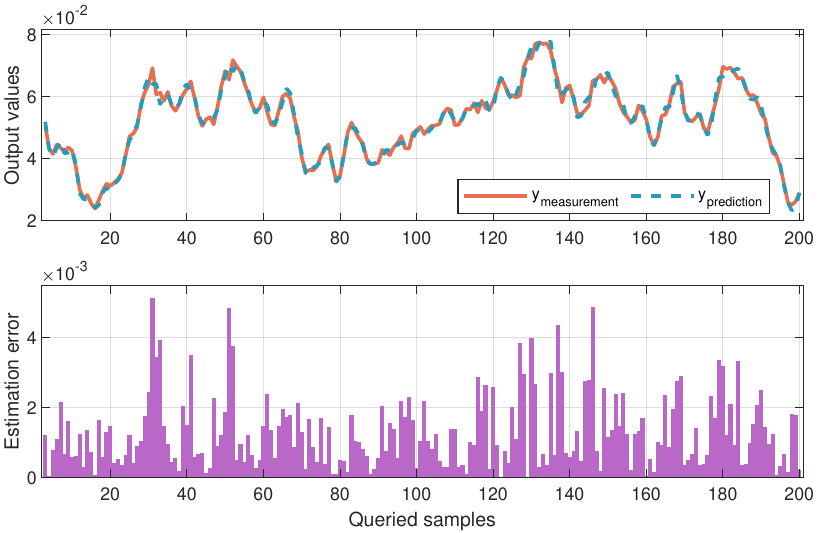}
	\caption{Two-tank benchmark, NARX-NN model: output predictions on test data after running Algorithm~\ref{alg:AL_NARX_ideal} for $N$ steps. Upper plot: first 200 output measurements $y_k$
    (\textcolor{burntSiennaRed}{orange} line) and predictions $\hat{y}_k = f(x_{k-1},\theta_N)$ (\textcolor{pacificBlue}{light blue} dashed line). Lower plot: output estimation errors ($P_0 = 10^{-2} I_{n_{\theta}}$, $Q = 10^{-10} I_{n_{\theta}}$, $R = 10^{-2} I_{n_{y}}$).
	}
	\label{fig: two_tank_test_outputs_NARX}
\end{figure}

\begin{figure}
	\centering
	\includegraphics[width=1.0\columnwidth]{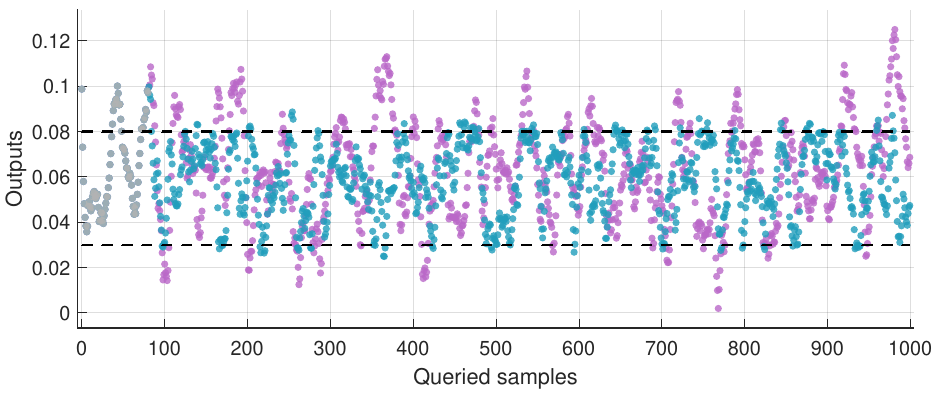}
	\caption{Two-tank benchmark, NARX-NN model, training dataset: Initial samples (\textcolor{Grey}{grey} dots), measurements $y_k$ without penalties (\textcolor{violetPurple}{purple} dots) and under soft constraints~\eqref{eq:RA-constraints} (\textcolor{pacificBlue}{light blue} dots), and constraints~\eqref{eq:TT-constraints} (black dashed lines). ($P_0 = 10^{-2} I_{n_{\theta}}$, $Q = 10^{-10} I_{n_{\theta}}$, $R = 10^{-2} I_{n_{y}}$). 
	}
	\label{fig: two_tank_const_outputs_NARX}
\end{figure}

Fig. \ref{fig: two_tank_test_outputs_NARX} shows the one-step-ahead predicted output $\hat{y}_k$ (upper plot) and its estimation error (lower plot) on the first 200 test data 
after running Algorithm~\ref{alg:AL_NARX_ideal} for $N$ steps. The model is able to predict the output $y_k$ accurately, with a small estimation error. Additionally, Fig. \ref{fig: two_tank_const_outputs_NARX} shows that introducing the constraints-violation penalty term in~\eqref{eq:soft-penalty} effectively prevents the outputs from violating the constraints~\eqref{eq:RA-constraints} during the active input design procedure. 

\begin{figure}
	\centering
	\includegraphics[width=1.0\columnwidth]{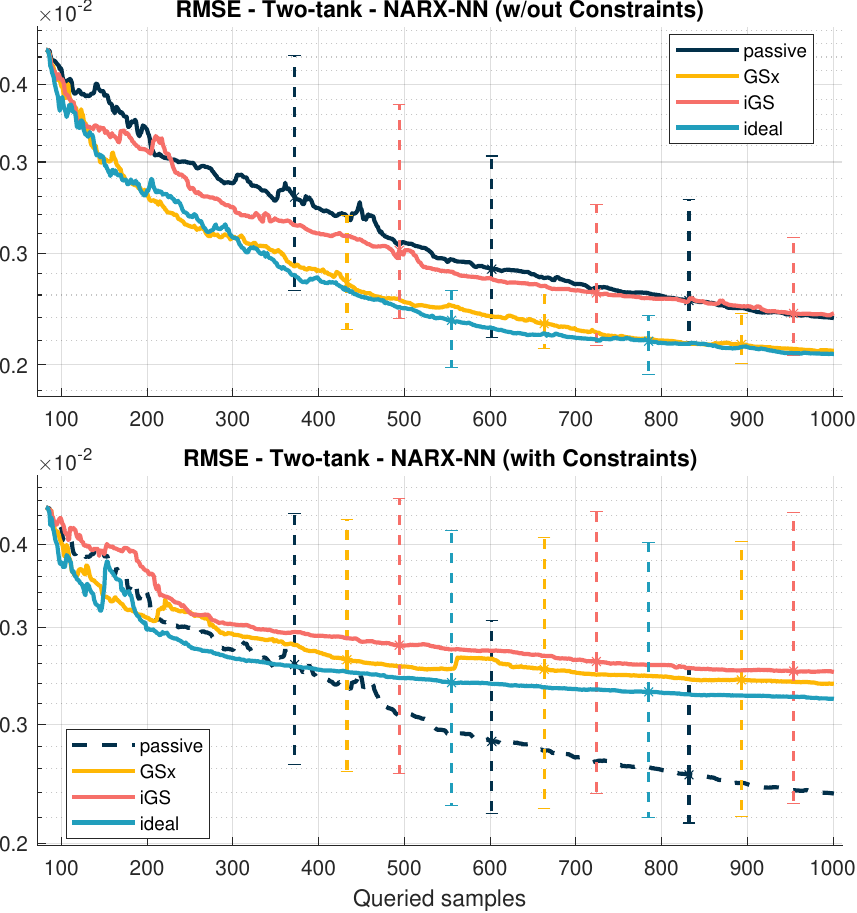}
	\caption{Two-tank benchmark, NARX-NN model: median RMSE without constraints (upper plot) and with constraints \eqref{eq:RA-constraints} (lower plot). Constraints are not taken into account and violated by passive sampling. Vertical lines denote mean absolute deviation ($P_0 = 10^{-2} I_{n_{\theta}}$, $Q = 10^{-10} I_{n_{\theta}}$, $R = 10^{-2} I_{n_{y}}$).}
	\label{fig: two_tank_OneStep_RMSE_NARX}
\end{figure}

We compare the performance of~\idealsysidnarx~when the AL method,~\ideal, is replaced by~\greedyx, and~\greedyxy~in Algorithm~\ref{alg:AL_NARX_ideal}. Fig. \ref{fig: two_tank_OneStep_RMSE_NARX} shows the RMSE and its mean absolute deviation of the predicted output $y_k$ on test data, with and without constraints, over 30 runs. For all the considered AL methods, the RMSE values are identical for $k \le N_i - 1$, indicating that they share the same initial randomly generated samples, resulting in the same parameters learned by the EKF. It is evident that \idealsysid~outperforms~\passive~and~\greedyxy~and is comparable to~\greedyx~in the unconstrained case (upper plot). With constraints considered, the penalty term limits the exploration of all AL methods, \idealsysid~is still better than~\greedyx~and~\greedyxy~
(lower plot). The~\passive~method achieves the best performance because output constraints are ignored and violated. This could be observed from the MCV values in Table~\ref{tab:MCV-TT}, which shows that the penalty term is effective in reducing constraint violations in all AL methods (the passive sampling method has more violations of the constraints). Additionally, it also shows that all AL methods better explore the output space than the passive method in the unconstrained case. 

EKF takes between $0.118 \ \rm ms$ and $0.562 \ \rm ms$ ($0.226 \ \rm ms$ on average) per sample 
to update the NARX-NN model parameters during one run of Algorithm~\ref{alg:AL_NARX_ideal}, which is negligible compared to the time required by the AL method to acquire a sample, see Table~\ref{tab:runtime-TT-NARX}. 
Generally, the execution time of~\ideal~is longer than that of~\greedyx~and~\greedyxy, since it requires computing the IDW variance and exploration functions. The execution time of all methods is proportional to the number of elements
in the pool $\UU$.

\begin{figure}
	\centering
	\includegraphics[width=1.0\columnwidth]{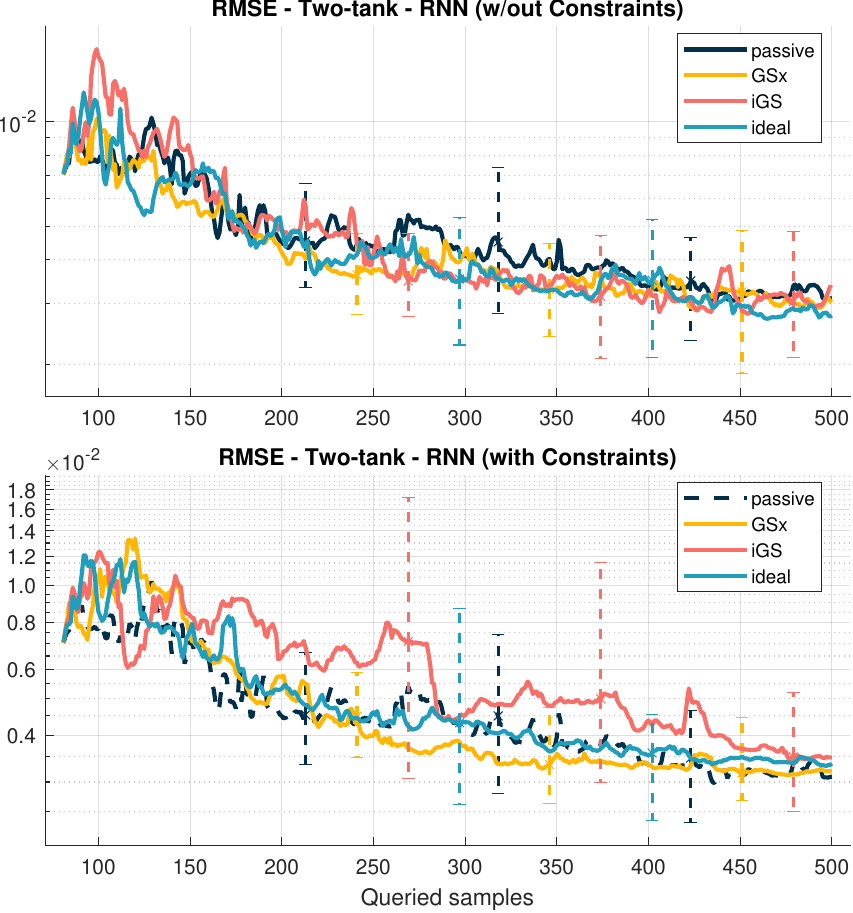}
	\caption{Two-tank benchmark, RNN model, median RMSE: no constraints (upper plot) and with constraints \eqref{eq:RA-constraints} (lower plot). Constraints are ignored during passive sampling. Vertical lines denote mean absolute deviation ($\bar P_0 = \operatorname{block diag}(4 \times 10^{-2} I_{n_x},\, 2 \times 10^{-1} I_{(n_{\theta_x} + n_{\theta_y})})$, $\bar Q_{x} = 10^{-8} I_{n_x}$, $\bar Q_{\theta} = 10^{-8} I_{(n_{\theta_x} + n_{\theta_y})}$, $\bar R = I_{n_{y}}$).}
	\label{fig: two_tank_OneStep_RMSE_RNN}
\end{figure}

As for the RNN model~\eqref{eq:RNN-f}, we set $n_x = 2$, $n_y = 1$, $n_u = 1$, $n_1^x = 8$, $n_2^x = 4$, and $n_1^y = 5$. The model is initially trained by L-BFGS-B in \texttt{jax-sysid}~\cite{Bem25} and then refined using EKF. The model parameters are updated recursively via EKF after measuring each new sample $y_k$. We set the~\idealsysidss~hyperparameters $\delta = 1000$, $\alpha = 0.001$, $N_i = 80$, $N = 500$, and $N_{\rm test} = 2000$, and apply Algorithm~\ref{alg:AL_SS_ideal} over 10 runs.
Fig. \ref{fig: two_tank_OneStep_RMSE_RNN} shows that~\ideal~outperforms passive sampling, similar to~\greedyx~and~\greedyxy~when there are no constraints (upper plot) while with constraints~\greedyx~achieves the best performance 
(lower plot). The \passive~method performs well in the constrained case by violating output constraints, 
as shown in the MCV values reported in Table~\ref{tab:MCV-TT}. 

The execution time of EKF to update the RNN model parameters ranges between $167 \ \rm ms$ and $1358 \ \rm ms$ ($253 \ \rm ms$ on average) per sample in one run of Algorithm~\ref{alg:AL_SS_ideal}, while reconstructing all past states by EKF and RTS smoothing ranges between $383 \ \rm ms$ and $1355 \ \rm ms$ ($432 \ \rm ms$ on average). 
Table~\ref{tab:runtime-TT-RNN} indicates that the average execution time increases progressively across the methods, following the order \passive, \greedyx, \greedyxy, and \ideal, as expected.

\begin{table}
    \centering
    \caption{Two-tank benchmark, NARX-NN model: MCV~\eqref{eq:MCV} on training data. All values are in $10^{-3}$ units ($y_{\rm min} = 0.03$, $y_{\rm max} = 0.08$).}
    \label{tab:MCV-TT}
    \setlength{\tabcolsep}{5.0pt}
    \renewcommand{\arraystretch}{1.2} 
    \begin{tabular}{llcccc}
        \hline
        \textbf{Model (runs)} & \textbf{Penalty} & \textbf{\passive} & \textbf{\greedyx} & \textbf{\greedyxy} & \textbf{\ideal} \\
        \hline
        NARX-NN (30) & w/out & 1.7 & 2.3 & 1.7 & 3.3 \\
        NARX-NN (30) & with    & 1.7 & 0.57 & 0.26 & 0.34 \\
		\hline
        RNN (10) & w/out & 1.5 & 1.5 & 2.9 & 1.6 \\
        RNN (10) & with    & 1.5 & 0.21 & 0.45 & 0.49 \\
        \hline
    \end{tabular}
\end{table}

\begin{table}
    \centering
    \caption{Two-tank benchmark, NARX-NN model: running time [ms] 
	\\(no penalty on output constraint violation).}
    \label{tab:runtime-TT-NARX}
    \setlength{\tabcolsep}{5.0pt}
    \renewcommand{\arraystretch}{1.2} 
    \begin{tabular}{lccc}
        \hline
        \textbf{Method} & \textbf{Min time} & \textbf{Avg time} & \textbf{Max time} \\
        \hline
        \passive   & 0.0161 & 0.0362 & 2.02 \\
        \greedyx   & 8.74   & 22.8   & 52.7 \\
        \greedyxy  & 9.57   & 25.2   & 61.5 \\
        \ideal     & 13.1   & 90.8   & 636  \\
        \hline
    \end{tabular}
\end{table}

To assess the quality of the identified model after $N$ samples are collected, we report the mean $R^2$-score on the test dataset in Table~\ref{tab:R2-TT}, averaged over $N_r$ runs. The results indicate that the learned NARX-NN and RNN models have a good accuracy. The \ideal~method consistently attains the highest $R^2$ values, followed by \greedyx, \greedyxy, and \passive, except in the constrained case, where \passive~performs comparably in spite of violating the output constraints. These findings are consistent with the RMSE results shown in Fig.~\ref{fig: two_tank_OneStep_RMSE_NARX} and~\ref{fig: two_tank_OneStep_RMSE_RNN} and the MCV values in Table~\ref{tab:MCV-TT}.

\begin{table}
    \centering
    \caption{Two-tank benchmark, RNN model: running time [ms] 
	\\(no penalty).} 
    \label{tab:runtime-TT-RNN}
    \setlength{\tabcolsep}{5.0pt}
    \renewcommand{\arraystretch}{1.2} 
    \begin{tabular}{lccc}
        \hline
        \textbf{Method} & \textbf{Min time} & \textbf{Avg time} & \textbf{Max time} \\
        \hline
        \passive   & 0.125 & 0.180 & 8.46 \\
        \greedyx   & 95.5  & 124   & 402  \\
        \greedyxy  & 104   & 146   & 575  \\
        \ideal     & 172   & 236   & 1352 \\
        \hline
    \end{tabular}
\end{table}

\begin{table}
    \centering
    \caption{Two-tank benchmark: mean $R^2$~\eqref{eq:R2} (\%) on test dataset}
    \label{tab:R2-TT}
    \setlength{\tabcolsep}{4.0pt}
    \renewcommand{\arraystretch}{1.2} 
    \begin{tabular}{llcccc}
        \hline
        \textbf{Model (runs)} & \textbf{Penalty} & \textbf{\passive} & \textbf{\greedyx} & \textbf{\greedyxy} & \textbf{\ideal} \\
        \hline
        NARX-NN (30) & w/out  & 97.92 & 98.17 & 97.90 & 98.17 \\
        NARX-NN (30) & with     & 97.92 & 97.05 & 96.91 & 97.08 \\
		\hline
        RNN (10) & w/out  & 96.67 & 96.64 & 95.50 & 97.24 \\
        RNN (10) & with     & 96.67 & 96.42 & 95.64 & 96.08 \\
        \hline
    \end{tabular}
\end{table}


\begin{figure}
	\centering
	\includegraphics[width=1.0\columnwidth]{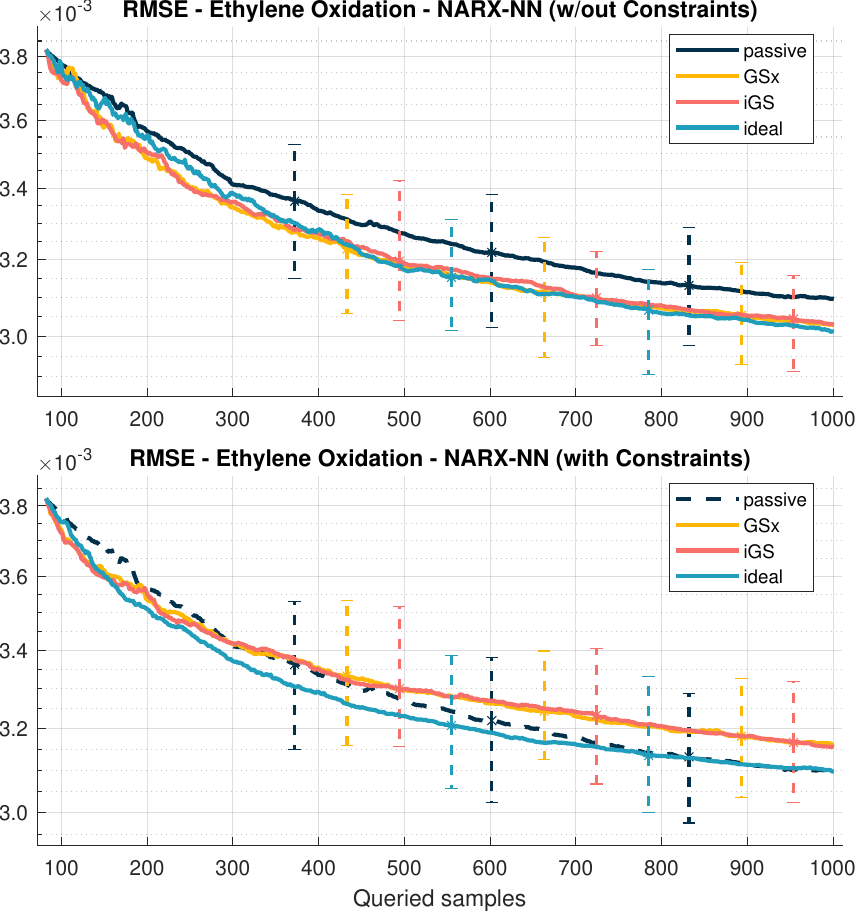}
	\caption{Ethylene oxidation benchmark predicted by a NARX-NN model, median RMSE: no constraints (upper plot) and with constraints \eqref{eq:EO-constraints} (lower plot). Constraints are ignored during passive sampling. Vertical lines denote mean absolute deviation ($P_0 = 10^{-2} I_{n_{\theta}}$, $Q = 10^{-10} I_{n_{\theta}}$, $R = 10^{-2} I_{n_{y}}$).}
	\label{fig: oxidation_OneStep_RMSE_NARX}
\end{figure}

\subsection{Ethylene oxidation benchmark}
We evaluate the proposed methods on the identification of an ethylene oxidation plant~\cite{DEC16}, a standard nonlinear MPC benchmark from the Model Predictive Control Toolbox for MATLAB. The plant features four states (gas density, C$_2$H$_4$ concentration, C$_2$H$_4$O concentration, and reactor temperature), one output (C$_2$H$_4$O concentration), and two inputs: the manipulated total volumetric feed flow rate $u$, and a measured disturbance $v$ (C$_2$H$_4$ feed concentration, fixed at its nominal value $0.5$). Data are generated by integrating the plant’s nonlinear ODEs with high accuracy, sampling every $T_s = 5$~s, and using initial conditions $x_0 = [0.9981,\, 0.4291,\, 0.0303,\, 1.0019]'$. Measurement noise $\eta_k = \gamma\, y_k\, \varepsilon_k$, with $\gamma = 0.08$, and $\varepsilon_k \sim \mathcal{N}(0,1)$ is injected on output data. The input $u$ is selected from the pool $\UU = \{0.0704,\, 0.0804,\, \ldots,\, 0.7042\}$ (step $0.01$). We also consider the output constraints
\begin{equation}
    0.02 \leq y_k \leq 0.05.
    \label{eq:EO-constraints}
\end{equation}

\begin{figure}
	\centering
	\includegraphics[width=1.0\columnwidth]{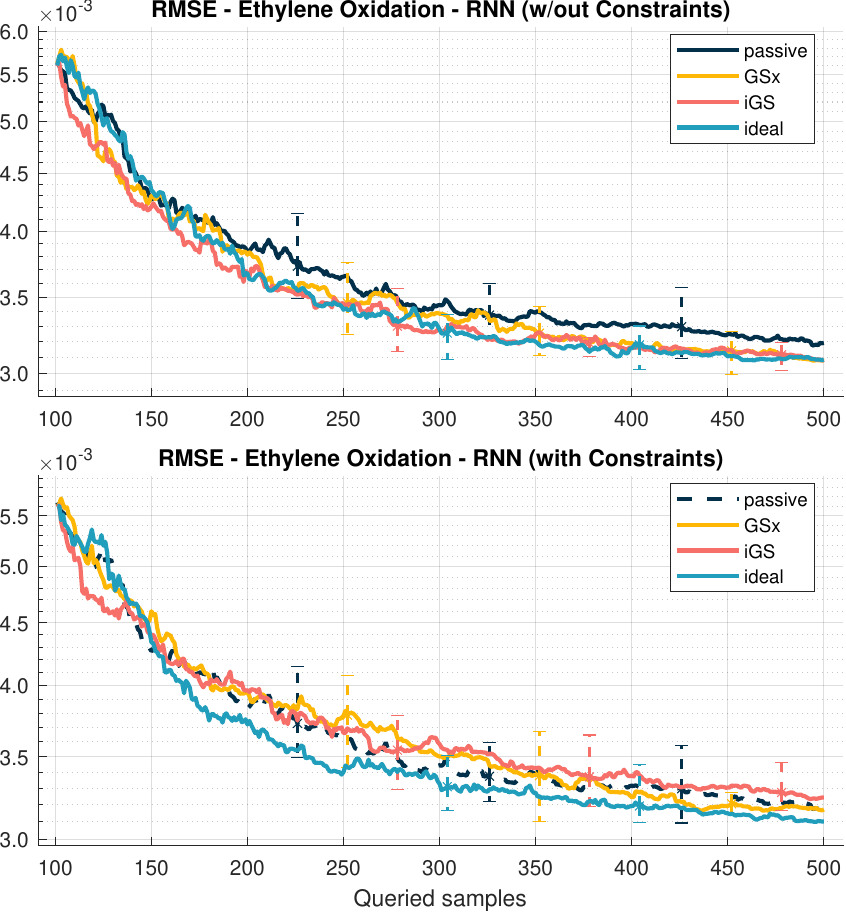}
	\caption{Ethylene oxidation benchmark predicted by a RNN model, median RMSE: no constraints (upper plot) and with constraints \eqref{eq:EO-constraints} ($\rho = 10^{6}$, lower plot). Constraints are ignored during passive sampling. Vertical lines denote mean absolute deviation ($\bar P_0 = 2I_{(n_x + n_{\theta_x} + n_{\theta_y})}$, $\bar Q_{x} = 10^{-10} I_{n_x}$, $\bar Q_{\theta} = 10^{-10} I_{(n_{\theta_x} + n_{\theta_y})}$, $\bar R = I_{n_{y}}$).}
	\label{fig: oxidation_OneStep_RMSE_RNN}
\end{figure}

For the NARX-NN model~\eqref{eq:NARX-f}, we use $n_1 = 8$, $n_2 = 6$, and set $\delta = 10$, $N_i = 80$, $N_{\rm test} = 1000$, and apply Algorithm~\ref{alg:AL_NARX_ideal}. A separate test set of $N_{\rm test} = 2000$ samples is generated with the same initial conditions and input pool. Fig. \ref{fig: oxidation_OneStep_RMSE_RNN} shows that all AL methods perform similarly and consistently outperform~\passive~in the non-constrained case (upper plot). With constraints considered, the penalty term restricts exploration, but \ideal~remains superior to \greedyx~and \greedyxy, and outperforms \passive~(lower plot). Note again that the good performance of the~\passive~method is because output constraints are ignored and violated, see the MCV values in Table~\ref{tab:MCV-EO}.

\begin{table}
    \centering
    \caption{Ethylene oxidation benchmark: MCV~\eqref{eq:MCV} on training dataset. All values are in $10^{-3}$ units. ($y_{\rm min} = 0.02$, $y_{\rm max} = 0.05$)}
    \label{tab:MCV-EO}
    \setlength{\tabcolsep}{5.0pt}
    \renewcommand{\arraystretch}{1.2} 
    \begin{tabular}{llcccc}
        \hline
        \textbf{Model (runs)} & \textbf{Penalty} & \textbf{\passive} & \textbf{\greedyx} & \textbf{\greedyxy} & \textbf{\ideal} \\
        \hline
        NARX-NN (30) & w/out & 0.71 & 1.4 & 2.1 & 2.7 \\
        NARX-NN (30) & with    & 0.71 & 0.33 & 0.28 & 0.59 \\
		\hline
        RNN (10) & w/out & 0.77 & 1.1 & 1.6 & 2.3 \\
        RNN (10) & with    & 0.77 & 0.41 & 0.42 & 0.64 \\
        \hline
    \end{tabular}
\end{table}

Then, for the RNN model~\eqref{eq:RNN-f}, we set $n_x = 4$, $n_y = 1$, $n_u = 1$, $n_1^x = 8$, $n_2^x = 6$, $n_1^y = 5$, and use $\delta = 100$, $\alpha = 1$, $N_i = 100$, $N = 500$, $N_{\rm test} = 2000$, and run Algorithm~\ref{alg:AL_SS_ideal} 10 times to collect statistics of the results.

Fig. \ref{fig: oxidation_OneStep_RMSE_RNN} shows that AL methods are comparable and outperform passive sampling (upper plot). When constraints are enforced, \ideal~gains the best performance (lower plot), where \passive~performs well at the expense of a violation of the constraints. This is reflected in the MCV values in Table~\ref{tab:MCV-EO}.

Table~\ref{tab:R2-EO} shows that both NARX-NN and RNN models achieve high $R^2$ values, though slightly lower than in the two-tank benchmark due to the higher noise level. The \ideal~method consistently yields the best $R^2$ performance, followed by \greedyx, \greedyxy, and \passive. In the constrained case, \passive~can appear competitive but it disregards output constraints, as also indicated by its higher MCV values.

\begin{table}
    \centering
    \caption{Ethylene oxidation benchmark: mean $R^2$ (\%) on test data}
    \label{tab:R2-EO}
    \setlength{\tabcolsep}{4.0pt}
    \renewcommand{\arraystretch}{1.2} 
    \begin{tabular}{llcccc}
        \hline
        \textbf{Model (runs)} & \textbf{Penalty} & \textbf{\passive} & \textbf{\greedyx} & \textbf{\greedyxy} & \textbf{\ideal} \\
        \hline
        NARX-NN (30) & w/out  & 92.57 & 92.88 & 92.87 & 92.96 \\
        NARX-NN (30) & with     & 92.57 & 92.24 & 92.28 & 92.56 \\
		\hline
        RNN (10) & w/out  & 92.22 & 92.79 & 92.81 & 92.78 \\
        RNN (10) & with     & 92.22 & 92.33 & 92.06 & 92.66 \\
        \hline
    \end{tabular}
\end{table}

\begin{table}
    \centering
    \caption{Robot arm benchmark: MCV~\eqref{eq:MCV} on training dataset. All values are in $10^{-2}$ units. ($y_{\rm min} = -0.4$, $y_{\rm max} = 0.4$)}
    \label{tab:MCV-RA}
    \setlength{\tabcolsep}{5.0pt}
    \renewcommand{\arraystretch}{1.2} 
    \begin{tabular}{llcccc}
        \hline
        \textbf{Model (runs)} & \textbf{Penalty} & \textbf{\passive} & \textbf{\greedyx} & \textbf{\greedyxy} & \textbf{\ideal} \\
        \hline
        NARX-NN (30) & w/out & 1.0 & 2.1 & 27 & 3.8 \\
        NARX-NN (30) & with    & 1.0 & 0.16 & 0.23 & 0.25 \\
		\hline
        RNN (10) & w/out 	   & 0.91 & 0.94 & 4.8 & 1.8 \\
        RNN (10) & with     & 0.91 & 0.22 & 0.25 & 0.29 \\
        \hline
    \end{tabular}
\end{table}

\begin{table}
    \centering
    \caption{Robot arm benchmark: mean $R^2$~\eqref{eq:R2} (\%) on test dataset}
    \label{tab:R2-RA}
    \setlength{\tabcolsep}{3.5pt}
    \renewcommand{\arraystretch}{1.2} 
    \begin{tabular}{llcccc}
        \hline
        \textbf{Model (runs)} & \textbf{Penalty} & \textbf{\passive} & \textbf{\greedyx} & \textbf{\greedyxy} & \textbf{\ideal} \\
        \hline
        NARX-NN (30) & w/out  & 98.16 & 98.55 & 94.74 & 98.62 \\
        NARX-NN (30) & with     & 98.16 & 97.35 & 97.40 & 97.54 \\
		\hline
        RNN (10) & w/out  & 92.04 & 93.22 & 89.06 & 95.79 \\
        RNN (10) & with     & 92.04 & 89.94 & 91.73 & 92.65 \\
        \hline
    \end{tabular}
\end{table}

\begin{figure}
	\centering
	\includegraphics[width=1.0\columnwidth]{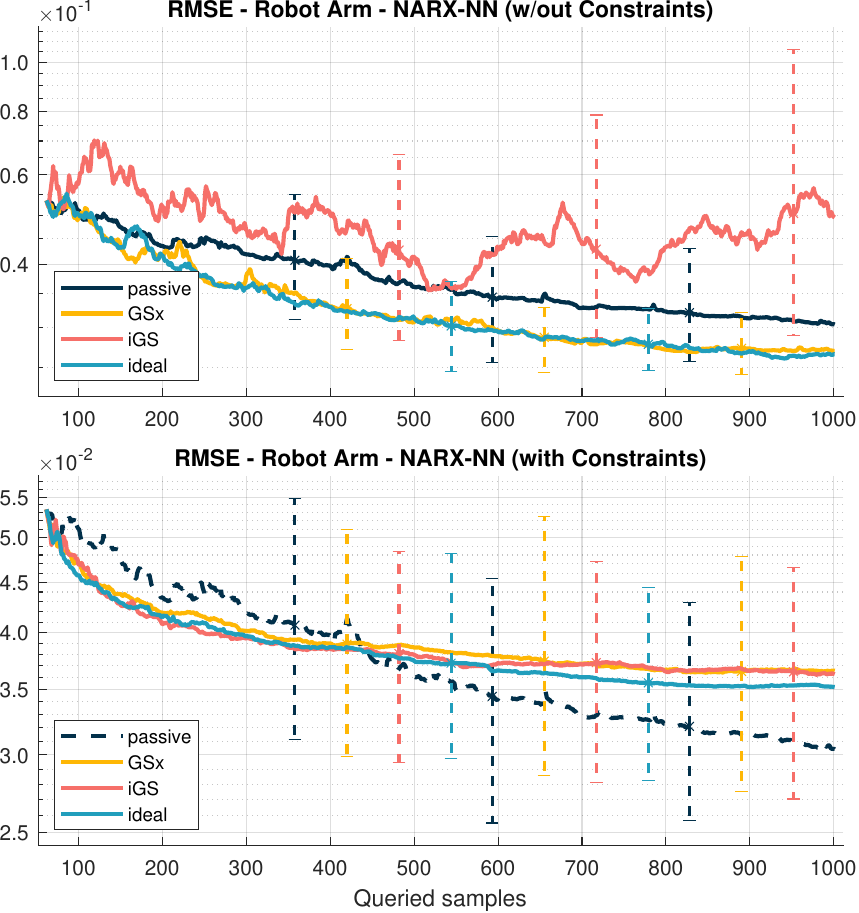}
	\caption{Robot arm benchmark predicted by a NARX-NN model, median RMSE: no constraints (upper plot) and with constraints \eqref{eq:RA-constraints} (lower plot). Constraints are ignored during passive sampling. Vertical lines denote mean absolute deviation ($P_0 = 10^{-2} I_{n_{\theta}}$, $Q = 10^{-10} I_{n_{\theta}}$, $R = 10^{-2} I_{n_{y}}$).}
	\label{fig: robot_arm_OneStep_RMSE_NARX}
\end{figure}

\subsection{Industrial robot arm benchmark}
We evaluate the methods on the industrial robot arm benchmark~\cite{WG06}, a nonlinear system identification problem from the System Identification Toolbox for MATLAB. The plant has five states (relative angular positions and velocities of the motor, gear-box, and arm), one output (motor angular velocity), and one input (applied motor torque). 
Data are generated by numerically integrating the plant's nonlinear ODEs using the adaptive stepsize Dormand-Prince Runge-Kutta method. The integration starts from the initial condition $x_0 = [0,\, 0,\, 0,\, 0,\, 0]'$, and the system output is sampled at intervals of $T_s = 0.0005$~s.
Measurement noise is Gaussian with standard deviation $0.08$ times the output value. The input pool is $\UU = \{-3.0,\, -2.99,\, \ldots,\, 3.0\}$ (step $0.01$). Output constraints are set as
\begin{equation}
    -0.4 \leq y_k \leq 0.4.
    \label{eq:RA-constraints}
\end{equation}

\begin{figure}
	\centering
	\includegraphics[width=1.0\columnwidth]{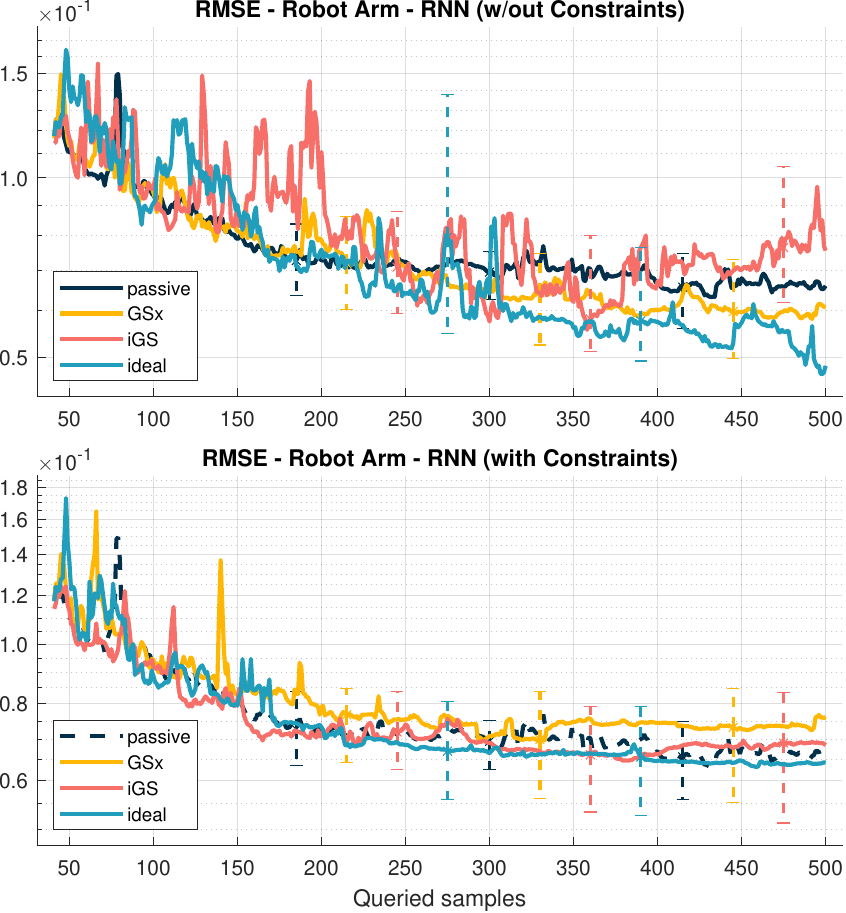}
	\caption{Robot arm benchmark predicted by a RNN model, median RMSE: no constraints (upper plot) and with constraints \eqref{eq:RA-constraints} (lower plot). Constraints are ignored during passive sampling. Vertical lines denote mean absolute deviation ($\bar P_0 = \operatorname{diag}(2 \times 10^{-1} I_{n_x},\, 2 \times 10^{-2} I_{(n_{\theta_x} + n_{\theta_y})})$, $\bar Q_{x} = 10^{-8} I_{n_x}$, $\bar Q_{\theta} = 10^{-8} I_{(n_{\theta_x} + n_{\theta_y})}$, $\bar R = 6.4 \times 10^{-3} I_{n_{y}}$).}
	\label{fig: robot_arm_OneStep_RMSE_RNN}
\end{figure}

The NARX-NN~\eqref{eq:NARX-f} model has $n_1 = 16$, $n_2 = 10$ hidden neurons. We set hyperparameters $\delta = 20$, $N_i = 60$, $N_{\rm test} = 1000$ and $N_{\rm test} = 1000$ in Algorithm~\ref{alg:AL_NARX_ideal}. Fig. \ref{fig: robot_arm_OneStep_RMSE_RNN} depicts that \idealsysidnarx~consistently outperforms \passive~and \greedyxy, and is comparable to \greedyx~in the unconstrained case. With exploration limited by penalty terms, \idealsysid~is still better than \greedyx~and~\greedyxy, and even outperforms \passive~before 400 samples are collected. MCV values are reported in Table~\ref{tab:MCV-RA}.

The RNN model~\eqref{eq:RNN-f} is configured with $n_x = 4$, $n_y = 1$, $n_u = 1$, $n_1^x = 8$, $n_2^x = 4$, $n_1^y = 5$. We use $\delta = 1000$, $\alpha = 0.1$, $N_i = 40$, $N = 500$, $N_{\rm test} = 2000$, and run Algorithm~\ref{alg:AL_SS_ideal} over 10 runs.
Fig. \ref{fig: robot_arm_OneStep_RMSE_RNN} shows that \idealsysidss~achieves the best overall performance compared to~\greedyx,~\greedyxy, and~\passive~in the unconstrained case. When constraints are enforced, \idealsysidss~continues to perform best. The good performance of the~\passive~method is due to ignoring output constraints, resulting in constraint violations, as reflected by the MCV values in Table~\ref{tab:MCV-RA}.

Table~\ref{tab:R2-RA} illustrates that \ideal~consistently yields the best $R^2$ performance, except in the constrained case, where, in spite of violating output constraints, \passive~performs comparably.

\section{Conclusion}\label{sec:cons}
We have presented and compared several active learning methods for online design of experiments tailored to the identification of nonlinear dynamical systems from input/output data, both in autoregressive and state-space form. By combining ideas of active learning for regression with recursive parameter estimation based on Kalman filtering, we demonstrated improved sample efficiency across a range of benchmark problems, as well as the ability of considering output constraints. All the proposed AL strategies provide benefits compared to standard random excitation. Among them, the \idealsysid~approach exhibited a robust and consistent performance. The results highlight the advantages of incorporating both exploration and exploitation terms in the acquisition function to ensure informative and diverse data collection. Future work will address the integration of these active learning strategies with closed-loop control, such as model predictive control. 

\section*{Acknowledgments}
The authors would like to thank Sampath Kumar Mulagaleti for fruitful discussions.

%
%
%


\bibliographystyle{plain}
\bibliography{online_active_learning_sysid_bib}

\end{document}